\documentclass{emulateapj}
\usepackage{amsmath}
\input{epsf}
\submitted{Accepted for publication in ApJ on May 24, 2016}

\usepackage{color}

\shorttitle{Tidally induced bars of galaxies in clusters}

\shortauthors{E. L. {\L}okas et al.}

\begin{document}

\title{Tidally induced bars of galaxies in clusters}

\author{Ewa L. {\L}okas\altaffilmark{1}, Ivana Ebrov\'{a}\altaffilmark{2}, Andr\'{e}s del Pino\altaffilmark{1},
Agnieszka Sybilska\altaffilmark{3,4}, E. Athanassoula\altaffilmark{5},\\ Marcin Semczuk\altaffilmark{1,6},
Grzegorz Gajda\altaffilmark{1} and Sylvain Fouquet\altaffilmark{1}}

\altaffiltext{1}{Nicolaus Copernicus Astronomical Center, Bartycka 18, 00-716 Warsaw, Poland}
\altaffiltext{2}{Astronomical Institute, The Czech Academy of Sciences, Bo\v{c}n\'{i} II 1401/1a,
CZ-141 00 Prague, Czech Republic}
\altaffiltext{3}{European Southern Observatory, Karl-Schwarzschild-Strasse 2, 85748 Garching bei M\"{u}nchen, Germany}
\altaffiltext{4}{Humboldt Fellow}
\altaffiltext{5}{Laboratoire d'Astrophysique de Marseille (LAM), UMR6110, CNRS/Universit\'e de Provence,
    Technop\^{o}le de Marseille Etoile, 38 rue Fr\'ed\'eric Joliot Curie, F-13388 Marseille C\'edex 13, France}
\altaffiltext{6}{Warsaw University Observatory, Al. Ujazdowskie 4, 00-478 Warsaw, Poland}

\begin{abstract}
Using $N$-body simulations we study the formation and evolution of tidally induced bars in disky galaxies in clusters.
Our progenitor is a massive, late-type galaxy similar to the Milky Way, composed of an exponential disk and an NFW dark
matter halo. We place the galaxy on four different orbits in a Virgo-like cluster and evolve it for 10 Gyr.
As a reference case we also evolve the same model in isolation. Tidally induced bars form on all orbits soon after the
first pericenter passage and survive until the end of the evolution. They appear earlier, are stronger, longer and have
lower pattern speeds for tighter orbits. Only for the tightest orbit the properties of the bar are controlled by the
orientation of the tidal torque from the cluster at pericenters. The mechanism behind the formation of the bars is the
angular momentum transfer from the galaxy stellar component to its halo. All bars undergo extended periods of buckling
instability that occur earlier and lead to more pronounced boxy/peanut shapes when the tidal forces are stronger. Using
all simulation outputs of galaxies at different evolutionary stages we construct a toy model of the galaxy population
in the cluster and measure the average bar strength and bar fraction as a function of clustercentric radius. Both are
found to be mildly decreasing functions of radius. We conclude that tidal forces can trigger bar formation in cluster
cores, but not in the outskirts, and thus cause larger concentrations of barred galaxies towards cluster center.
\end{abstract}


\section{Introduction}

Barred galaxies constitute between $\sim$30 and 70\% of the galaxy population, depending on the exact
definition of the bar, morphological type, environment etc.
(Aguerri et al. 2009; Buta et al. 2010, 2015; Cheung et al. 2013; and references therein). Bars seem to be
important for the internal evolution of galaxies, e.g. since they provide means of transport for gas, stars as
well as redistribution of angular momentum, and so are able to profoundly influence the structure of host objects (e.g.
Athanassoula 2013). As a result, bars are expected to be the driving force behind the formation of
inner galactic structures, such as star-forming rings (Buta et al. 2004), and they also contribute to the build-up of
bulge-like structures, the so-called pseudobulges (e.g. Fathi \& Peletier 2003; Chung \& Bureau 2004; Kormendy \&
Kennicutt 2004; Athanassoula 2005).

With the advent of large, often high-resolution, photometric surveys like the
Spitzer Survey of Stellar Structure in Galaxies ($\mathrm{S^4G}$, Sheth et al. 2010; Kim et al. 2015) or
Calar Alto Legacy Integral Field Area (CALIFA) Survey (S\'{a}nchez et al. 2012)
it became possible to not only study the properties of individual barred galaxies in much more detail but also assess
the statistical properties of barred galaxies.
Observational studies of barred galaxies focus either on a large number of objects to obtain statistical
information on their properties across a number of parameters, or they can be detailed
studies of a smaller number of objects (e.g. Aguerri et al. 2015; Seidel et al. 2015) or even single galaxies
(e.g. Gadotti et al. 2015). The
former are useful e.g. for comparison with high-redshift objects, while the latter can be used to perform detailed
comparisons with simulations.

Numerical simulations have shown that bars can form spontaneously from dynamical instabilities
inherent in self-gravitating axisymmetric disks
(e.g. Efstathiou et al. 1982; for a review see Athanassoula 2013).
Purely self-gravitating disks of early simulations were instantly susceptible to bar formation, with
added spherical potential delaying but not preventing it (Ostriker \& Peebles 1973; Athanassoula 2002).
Bars were also found to grow slower in hotter disks i.e. those characterized by higher velocity dispersions
(Athanassoula \& Sellwood 1986; Athanassoula 2003).
Bar growth (in mass, length as well as strength) is predicted to be due to the transformation of stellar
orbits in the disk from circular to elongated ones as a result of angular momentum loss of the
involved stars. At the same time as the bar strength increases, its pattern speed is expected to decrease
(e.g. Athanassoula 2003; Villa-Vargas et al. 2009).
An important episode in the secular evolution of a bar is that of buckling
instability leading to the formation of a boxy/peanut shape (Combes et al. 1990; Raha et al. 1991)
and usually occurring about 1-2 Gyr after the formation of the bar (Martinez-Valpuesta et al. 2006).

Another channel for the formation of bars in galaxies may be related to interaction with other galaxies of
similar, larger or smaller size. It has been demonstrated that interactions with a perturber can lead to the
formation of tidally induced bar in a galaxy (e.g. Gerin et al. 1990; Noguchi 1996; Miwa \& Noguchi 1998;
Mayer \& Wadsley 2004; Berentzen et al. 2004; Lang et al. 2014; Goz et al. 2015).
Recently, {\L}okas et al. (2014) investigated the evolution of a tidally induced bar in a dwarf
galaxy orbiting a Milky Way-like host, a good candidate for such a bar being the Sagittarius dwarf
({\L}okas et al. 2010). Tidally induced bars could also form as a result of interaction of normal-size or dwarf
galaxies with a cluster-like potential (Byrd \& Valtonen 1990; Mastropietro et al. 2005). In particular,
Mastropietro et al. (2005) have shown that bars are an intermediate stage in the evolution of low-mass galaxies under
strong cluster tidal fields from late to early type objects. If such a mechanism was indeed dominant
in clusters we should observe an increased fraction of barred galaxies towards the cluster center.

Unfortunately, there is at present no clear evidence for such a relation, although early studies seemed to support
this hypothesis. An analysis of the clustercentric distances of barred galaxies in the Coma cluster by
Thompson (1981) showed that a significantly larger fraction of barred galaxies are located at the cluster core
than at larger clustercentric distances. In a study of the Virgo cluster,
Andersen (1996) found that the barred disk galaxies are more centrally concentrated than the unbarred disks.
More recently, Barazza et al. (2009) studied the bar fraction in clusters at moderate redshifts $z=0.4-0.8$
and found that the bar fraction decreases strongly with distance if normalized by the cluster virial radius.

However, other studies seem to have reached less firm or even conflicting conclusions.
For example, M\'{e}ndez-Abreu et al. (2010) studied the bar fraction in the Coma cluster (based on a sample of 190
galaxies spanning a 9 magnitude range) and found no dependence on clustercentric distance, suggesting that the
environment is not the most important factor and plays a secondary role in bar formation and/or evolution.
Lansbury et al. (2014) analyzed bars in S0 galaxies in Coma (using SDSS DR8 data) and found an increase
in the bar fraction towards the cluster core, but at a low significance level.
In another study, Cervantes Sodi et al. (2015) find (from the analysis of SDSS DR7 data) that the bar
fraction is not directly dependent on the group/cluster environment, but is a strong function of stellar mass.

In general, external factors in bar formation are difficult to quantify. In order to aid in such endeavor, in this
paper we study in detail the tidal evolution of a massive late-type disky galaxy similar to the Milky Way.
By evolving the galaxy in a Virgo-like cluster for 10 Gyr on four distinct orbits, we investigate
the influence of tidal forces of varying strength on the bar formation and evolution. As a reference case,
we also analyze bar properties in the same model evolved in isolation.
Note that our purpose here is to investigate the effect of the tidal force of the cluster as a whole
and not the effect of the short term encounters between individual galaxies that may also influence the formation
and evolution of bars in cluster galaxies. However, this second process is expected to be less important for
normal-size galaxies we consider here than for dwarfs or low surface brightness objects (Moore et al. 1999).

The paper is organized as follows. In section 2 we describe the simulations used for this study. Section 3 presents
a general description of the evolution of the galaxies in the cluster including the main characteristics such
as kinematics and shape. Later, in section 4 we investigate in more detail the properties of
the bars formed during the evolution: their
surface density distribution, bar mode strength and pattern speed as well as buckling instability.
Finally, in section 5
we use a few hundred simulation snapshots at varying clustercentric radii sampled from four simulations
to create a toy model of the cluster galaxy population in which we measure the expected fraction of barred galaxies as a
function of radius and attempt a comparison with observations. The discussion follows in section 6.

\section{The simulations}

The initial conditions for our simulations consisted of $N$-body realizations of the Virgo cluster and a Milky
Way-like progenitor generated via procedures described in Widrow \& Dubinski (2005)
and Widrow et al. (2008). The procedures allow for the creation of $N$-body models of galaxies and halos very near
equilibrium. The Virgo cluster was approximated as a single-component (Navarro et al. 1997, NFW)
spherical dark matter halo of $10^6$ particles with parameters estimated
by McLaughlin (1999) from the combined analysis of X-ray and kinematic data and renormalized by Comerford \&
Natarajan (2007) to match our definitions of the virial mass and concentration (following {\L}okas \& Mamon 2001).
Namely, our Virgo cluster has the virial
mass $M_{\rm C} = 5.4 \times 10^{14}$ M$_{\odot}$, the concentration $c=3.8$ and an isotropic velocity distribution.
In order to make the cluster mass finite we introduce a smooth cut-off in the density profile at the scale of the
virial radius (2.1 Mpc) so that the total mass is equal to the virial mass.

We compared the line-of-sight velocity dispersion profile of such an $N$-body realization to the
presently available kinematic data for galaxies in the region of Virgo from the NED database. We find that the
velocity dispersion profile of the generated model (although based on parameters aiming to reproduce the present
properties of the cluster) falls significantly below the presently measured one. This means that
our model may be considered a good approximation for the average gravitational potential in the Virgo cluster over
the last few Gyr rather than just the present-day maximum value. Once the evolution of mass content of clusters (and
in Virgo in particular) is known in more detail, the results presented here can be mapped to a given epoch in
the evolution of the cluster. Note also, that the Virgo cluster is not a very massive cluster and we may expect
typical, more massive clusters to be even more effective in inducing bars since their tidal forces will be stronger.

As our model of the progenitor galaxy we adopted a rather large, late-type galaxy,
similar to the Milky Way. We have used a model close to the model MWb of Widrow \& Dubinski (2005), with just
two components: an NFW dark halo and an exponential disk, but no classical bulge. Each of the two components was
made of $10^6$ particles. The dark matter halo had a virial mass
$M_{\rm H} = 7.7 \times 10^{11}$ M$_{\odot}$ and concentration $c=27$. The disk had a mass $M_{\rm D}
= 3.4 \times 10^{10}$ M$_{\odot}$, the scale-length $R_{\rm D} = 2.82$ kpc and thickness $z_{\rm D} = 0.44$ kpc.
Both components were smoothly cut off at appropriate scales.  The minimum value of the
Toomre parameter of this realization was
$Q=2.1$ so we expect the model to be stable against bar formation for at least a few Gyr.

We place the progenitor galaxy on a few typical, eccentric orbits in the Virgo cluster with apo- to pericenter
distance ratio $D_{\rm apo}/D_{\rm peri}
=5$ (Ghigna et al. 1998). The values of the apo- and pericenters were chosen to cover a wide range of distances where
most of the galaxies in Virgo are observed. Our tightest orbit with $D_{\rm peri} =0.1$ Mpc
is the smallest possible, that can still be considered unaffected by the central galaxy, while our most extended orbit
reaches the outskirts of the cluster. In all simulations the disk of the progenitor was coplanar and
exactly prograde with the orbit.
We performed four such simulations, which will be referred to as S1-S4, with identical initial
conditions except for the orbit sizes. As a reference case we have also evolved our progenitor galaxy in isolation
and will refer to this simulation as S5. The orbital parameters of the simulations are summarized in
Table~\ref{initial}. The second and third column of the Table list the apo- and pericentric distances, the fourth gives
the radial orbital period, the fifth the number of pericenter passages occurring within the time during which we followed
the evolution and the last one the color coding to be used throughout the paper.

The evolution of the system in each simulation was followed for 10 Gyr using the GADGET-2 $N$-body code
(Springel et al. 2001; Springel 2005) with outputs saved every 0.05 Gyr.
The adopted softening scales were $\epsilon_{\rm D} = 0.1$ kpc and
$\epsilon_{\rm H} = 0.7$ kpc for the disk and halo of the galaxy while
$\epsilon_{\rm C} = 14$ kpc for the halo of the Virgo cluster, respectively. This large softening of the host
(of the order of 5 scale-lengths of the galaxy disk)
was set so as to minimize the effect of the rather large mass of the cluster particles on the evolution.

\begin{table}
\begin{center}
\caption{Orbital parameters of the simulations}
\begin{tabular}{lccccl}
\hline
\hline
Simulation    &  $D_{\rm apo}$ & $D_{\rm peri}$ & $T_{\rm orb}$ & $n_{\rm peri}$ & Line color \\
              &   [Mpc]    &    [Mpc]       &   [Gyr]    &     &       \\
\hline
\ \ \ \ S1    &\ 0.5       &\   0.1           &    1.3 &  9      & \ \ red    \\

\ \ \ \ S2    &\ \, 0.75   &\ \, 0.15         &    1.9 &  6      & \ \ green    \\

\ \ \ \ S3    &\ 1.0       &\   0.2           &    2.5 &  4      & \ \ cyan  \\

\ \ \ \ S4    &\ 1.5       &\   0.3           &    3.7 &  3      & \ \ blue   \\

\ \ \ \ S5    &\ --        &\   --            &    --  &  --     & \ \ black   \\

\hline
\label{initial}
\end{tabular}
\end{center}
\end{table}

\begin{figure}
\begin{center}
    \leavevmode
    \epsfxsize=8cm
    \epsfbox[0 10 189 185]{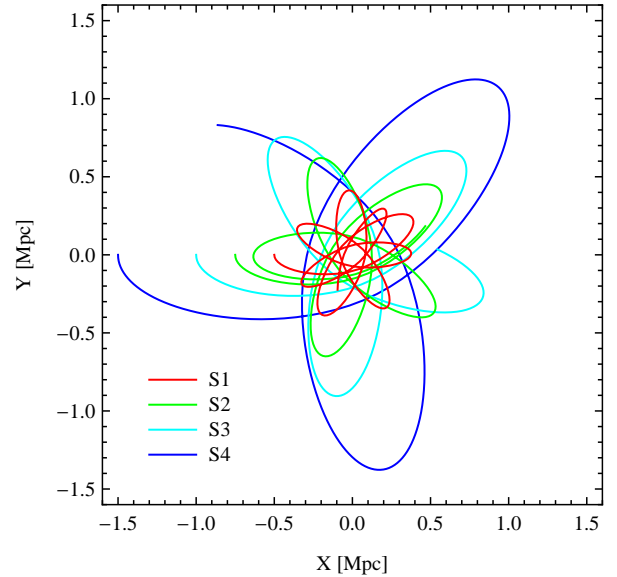}
\end{center}
\caption{The orbits of the simulated galaxies in the cluster in projection onto the orbital plane.
The orbits in simulations S1-S4 are shown with the red, green, cyan and blue line respectively. The same
color coding is used in the following figures.}
\label{orbitxy1234}
\end{figure}

\begin{figure}
\begin{center}
    \leavevmode
    \epsfxsize=8cm
    \epsfbox[0 10 186 270]{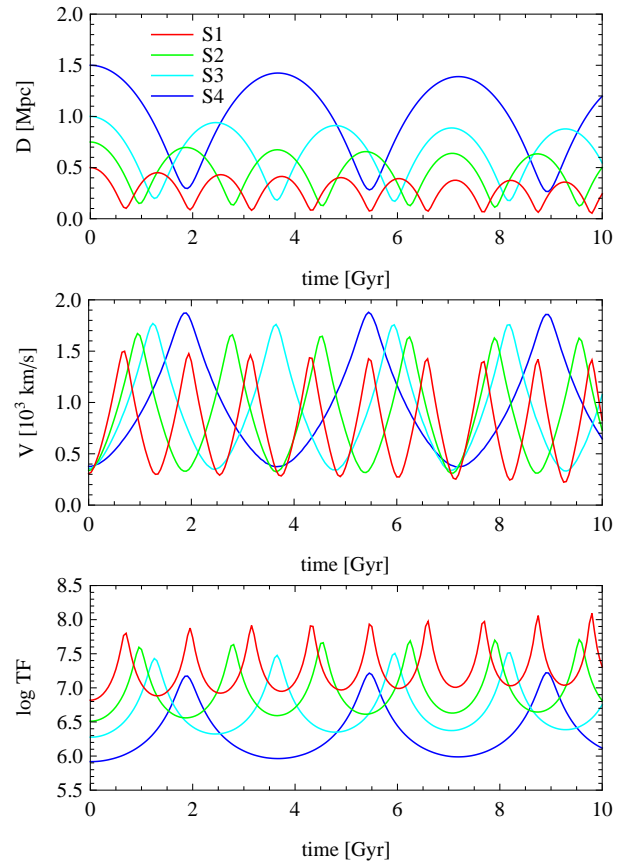}
\end{center}
\caption{The evolution of the clustercentric distance (upper panel) and the orbital velocity (middle panel)
of the galaxies in the cluster. The lower panel shows the evolution of the tidal force (TF) experienced by the galaxies.}
\label{orbit1234}
\end{figure}

\section{Evolution of the galaxies}

\subsection{Orbital evolution}

The orbital evolution of the galaxies in the cluster during 10 Gyr
is illustrated in Figures~\ref{orbitxy1234} and \ref{orbit1234}.
In each simulation the initial position of the progenitor was at the coordinates $(X,Y,Z) = (-D_{\rm apo},0,0)$ kpc
of the simulation box and the velocity vector of the galaxy was toward the negative $Y$ direction.
Figure~\ref{orbitxy1234}
shows the orbits of the simulated galaxies in projection onto the orbital plane $XY$.
Figure~\ref{orbit1234} compares the distance (upper panel) from the cluster center and the
orbital velocity (middle panel) of the galaxies in the cluster. The radial orbital periods for the different
simulations (between the first two apocenters) are listed in the fourth column of Table~\ref{initial}.
The galaxies experience a different number
of pericenter passages: nine, six, four and three respectively for S1 to S4. Note that some orbital decay is seen in the
orbits as a result of dynamical friction so the apo- and pericenter distances as well as orbital periods decrease in time.

The lower panel of Figure~\ref{orbit1234} shows the logarithm of the tidal force experienced by the stars
of the progenitor
galaxy approximated as $r M(<D)/D^3$ (and expressed in $M_{\odot}$ kpc$^{-2}$)
where $r$ is the distance from the progenitor center and $M(<D)$ is the mass of the
cluster contained within the galaxy's distance from the cluster center, $D$, as it moves on its orbit. For the calculation,
we adopted $r=7$ kpc, a scale-length found below to correspond to the typical length of the bar. We note that although the
mass of the cluster contained within $D$ is significantly smaller for the tighter orbit, the $D^{-3}$ dependence of the
tidal force on distance prevails and the tidal force is systematically stronger for tighter orbits than for more extended
ones both at pericenters and apocenters.

The values of log TF at the first pericenter passage vary from 7.2 for simulation S4 to 7.8 for S1. These values
can be compared with those characteristic for the configurations applied in the study of Miwa \& Noguchi (1998). They
used close to point-mass perturbers of 1 and 3 times the mass of the perturbed galaxy and pericenter distances of 40 kpc.
These values translate to log TF of 7.3 and 7.8 in our units. Miwa \& Noguchi concluded that these two values bracket
the transition between two regimes of tidal bar formation: when the tidal perturbation is relatively weak
the bar properties are determined mostly by the internal structure
of the perturbed galaxy while a sufficiently strong tidal perturbation
washes out the intrinsic structure of the target galaxy and creates a bar with properties determined by
the parameters of the tidal encounter. Comparing these values with ours, we may expect that in simulation S4, where the
tidal force is weak and below the range, the bar will be very weakly affected by tides while for the tighter orbits S1-S3
we should see bars with different properties.

\begin{figure}
\begin{center}
    \leavevmode
    \epsfxsize=8.2cm
    \epsfbox[0 10 186 261]{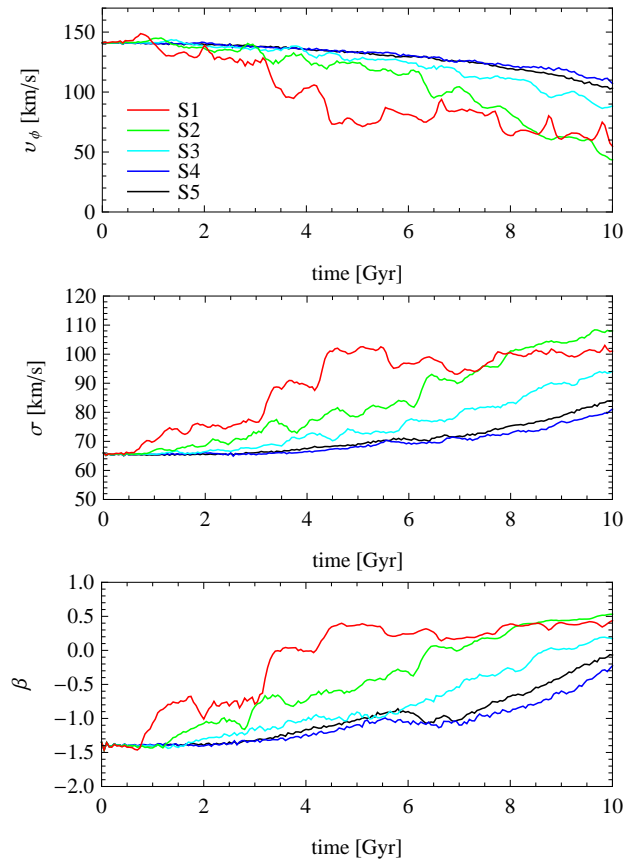}
\end{center}
\caption{Evolution of the kinematic properties of the stellar component of the galaxies within 7 kpc in time.
The upper panel shows the mean rotation
velocity around the shortest axis $v_{\phi}$ and the second panel
plots the 1D velocity dispersion averaged from three dispersions measured
in spherical coordinates. The last panel illustrates the evolution of the anisotropy parameter $\beta$.}
\label{kinematics7}
\end{figure}

\begin{figure}
\begin{center}
    \leavevmode
    \epsfxsize=8cm
    \epsfbox[0 9 186 267]{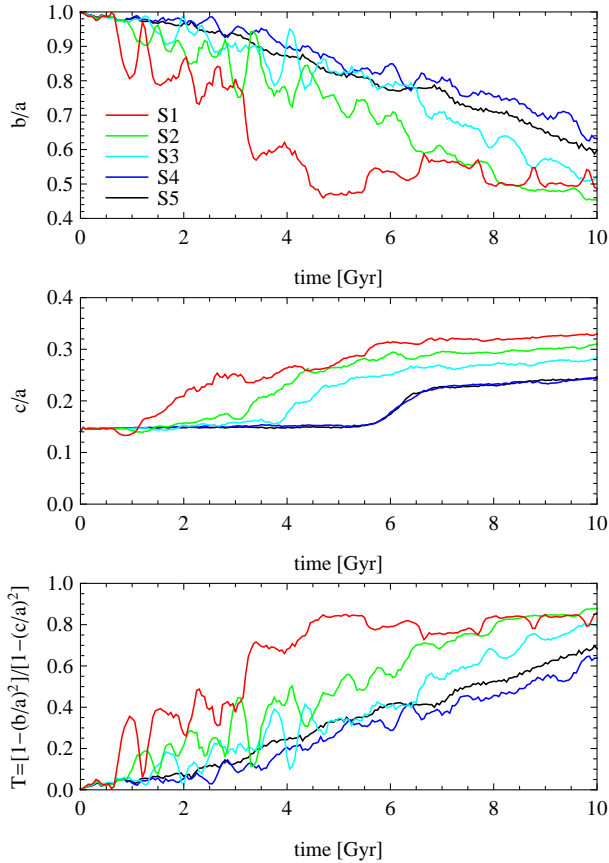}
\end{center}
\caption{Evolution of the shape of the stellar component of the
galaxies within 7 kpc in time. The three panels from top to bottom show
respectively the values of the ratio of the intermediate to longest axis $b/a$, the shortest to longest axis $c/a$,
and the triaxiality parameter $T$.}
\label{shape7}
\end{figure}

\subsection{Evolution of kinematics}

In order to measure the global kinematic properties of the stellar component of the galaxies as a function of time we
calculated the direction of the principal axes
of the inertia tensor of the stars within 7 kpc from the galactic center and rotated the stellar distribution
to align the coordinate system with the principal axes so that the $x$ coordinate is along the major axis, $y$ along
the intermediate, and $z$ along the shortest one. The choice of the
radius of 7 kpc, corresponding to about $2.5 R_{\rm D}$, was motivated by the typical length of the bars in
the later stages of the evolution and
will be justified below, when we discuss the properties of bars in more detail. We then introduce a standard system of
spherical coordinates and calculate the mean velocities and dispersions of the stars (see {\L}okas et al. 2014 for
details).

Figure~\ref{kinematics7} illustrates the kinematic properties of the stellar components as a function of time for
the five simulations. The upper panel shows the rotation of the stars around the minor axis $v_{\phi}$.
As expected from previous studies
of tidal stirring of galaxies, in general the rotation decreases with time, i.e. the streaming motion of the stars
is replaced
by random motions, as confirmed by the increasing 1D velocity dispersion of the stars seen in the second panel.
During the first 5 Gyr of evolution the dependence on the orbit is clearly monotonic: the effect is strongest for the
tightest orbit (S1) and weakest for the most extended one (S4).
After that the trend of decreasing rotation and increasing dispersion
is not so clearly present for S1. The rotation can both decrease and increase at a given
pericenter passage depending on the orientation of the bar at this moment. This is the same effect as was discussed
for tidally induced bars in dwarf galaxies by {\L}okas et al. (2014, see their figure 11): the bar is speeded up
if the tidal torque acts in the same direction as the bar rotates,
while it is slowed down if the torque acts in the opposite
direction. As a result, at the end of the evolution S2 is more evolved than S1 in a sense of having lower rotation
and higher velocity dispersion. This is just due to the more `favorable' orientation of its bar at pericenter
passages. Let us also note that the evolution on the most extended orbit (S4) is very similar (even slightly slower)
to the case of the galaxy evolving in isolation (S5). This means that at this most extended orbit the tidal force
is too weak to significantly affect the evolution.

\begin{figure*}
\begin{center}
    \leavevmode
    \epsfxsize=4.22cm
    \epsfbox[0 0 185 200]{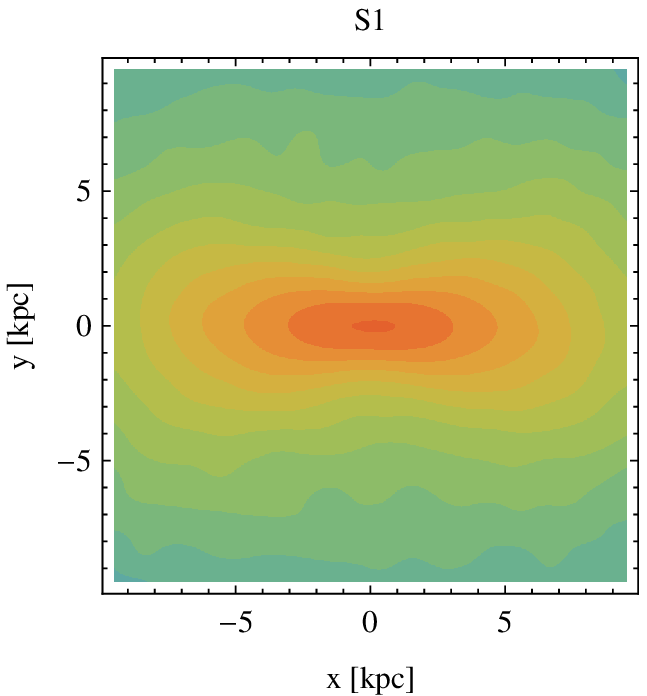}
\leavevmode
    \epsfxsize=4.22cm
    \epsfbox[0 0 185 200]{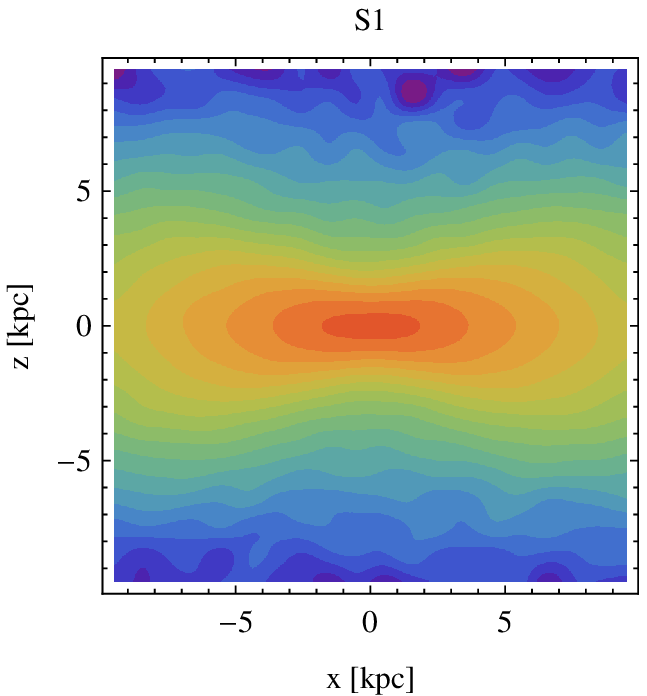}
\leavevmode
    \epsfxsize=4.22cm
    \epsfbox[0 0 185 200]{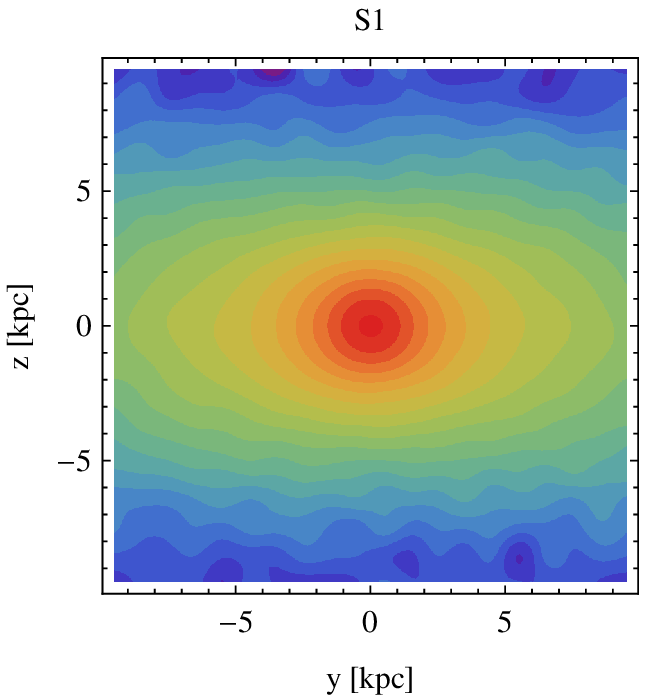}
\leavevmode
    \epsfxsize=0.805cm
    \epsfbox[53 -28 87 149]{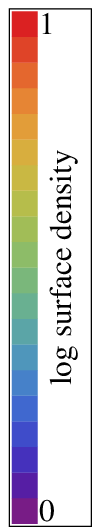}
\leavevmode
    \epsfxsize=4.22cm
    \epsfbox[0 0 185 200]{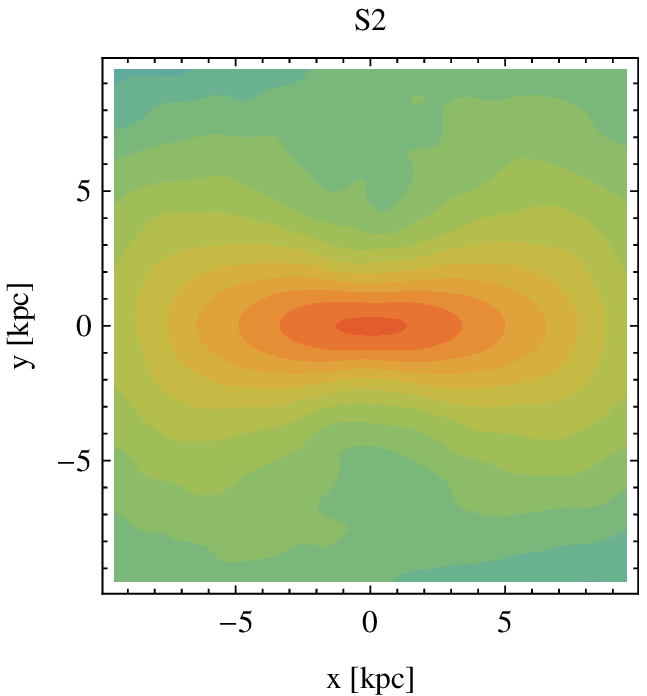}
\leavevmode
    \epsfxsize=4.22cm
    \epsfbox[0 0 185 200]{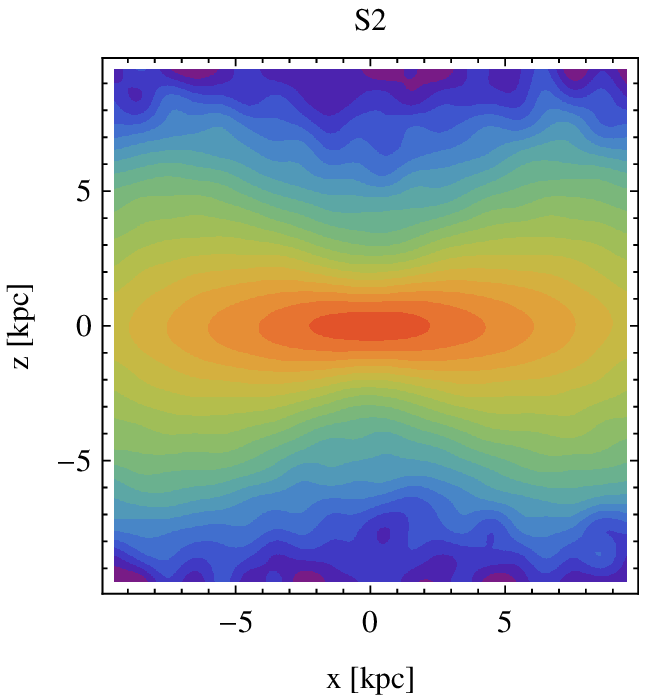}
\leavevmode
    \epsfxsize=4.22cm
    \epsfbox[0 0 185 200]{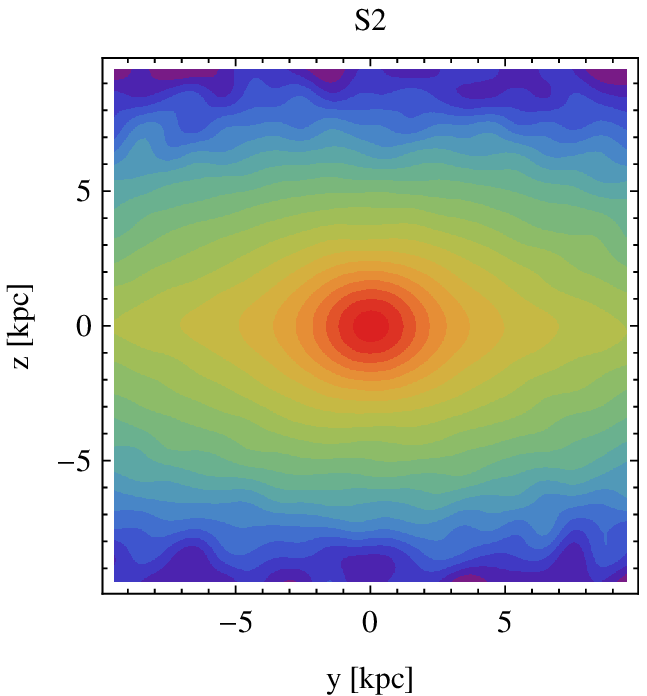}
\leavevmode
    \epsfxsize=0.805cm
    \epsfbox[53 -28 87 149]{legend.eps}
\leavevmode
    \epsfxsize=4.22cm
    \epsfbox[0 0 185 200]{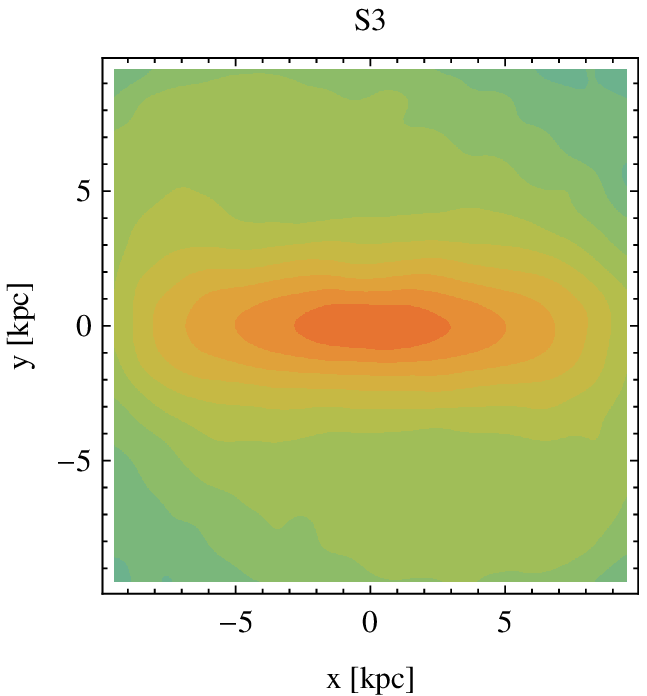}
\leavevmode
    \epsfxsize=4.22cm
    \epsfbox[0 0 185 200]{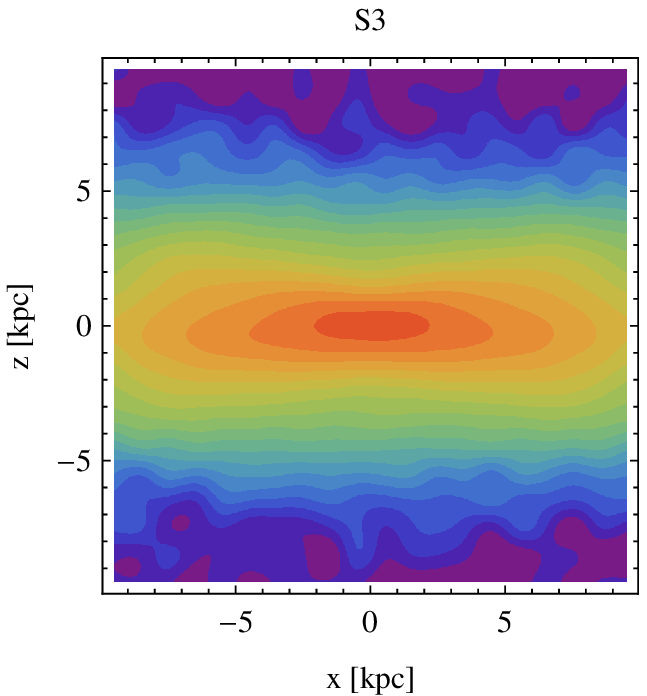}
\leavevmode
    \epsfxsize=4.22cm
    \epsfbox[0 0 185 200]{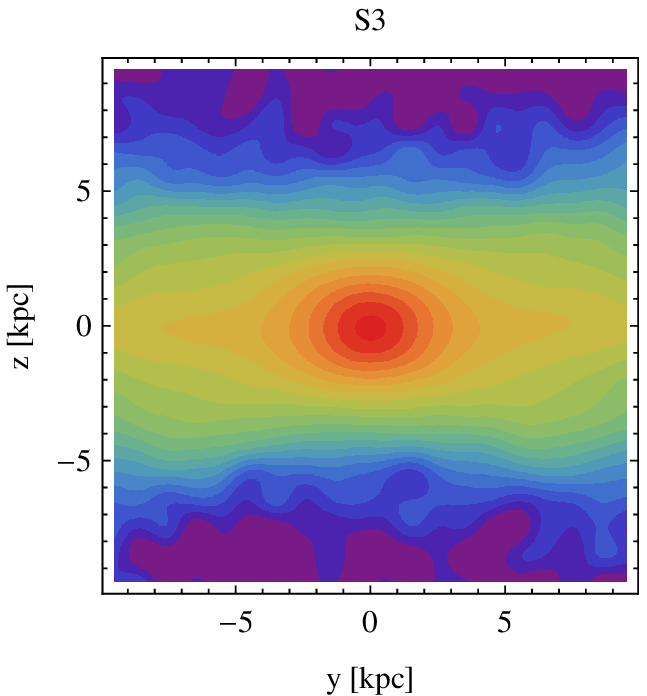}
\leavevmode
    \epsfxsize=0.805cm
    \epsfbox[53 -28 87 149]{legend.eps}
\leavevmode
    \epsfxsize=4.22cm
    \epsfbox[0 0 185 200]{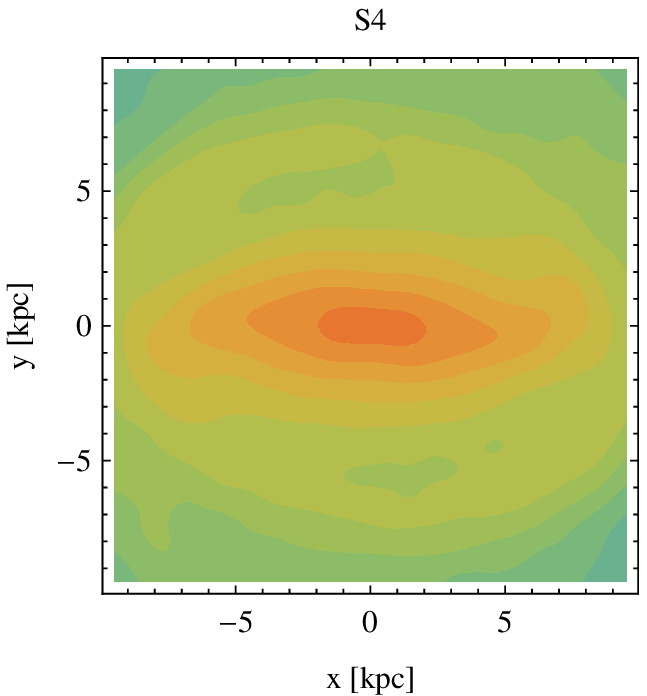}
\leavevmode
    \epsfxsize=4.22cm
    \epsfbox[0 0 185 200]{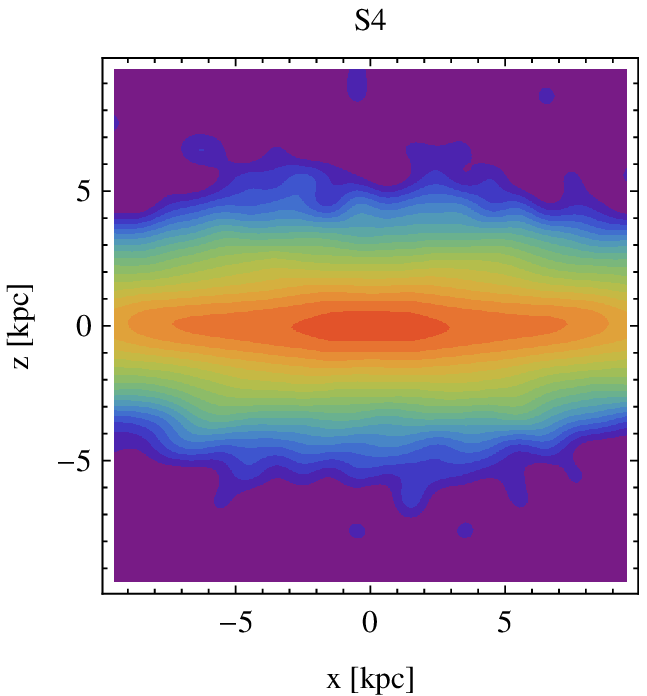}
\leavevmode
    \epsfxsize=4.22cm
    \epsfbox[0 0 185 200]{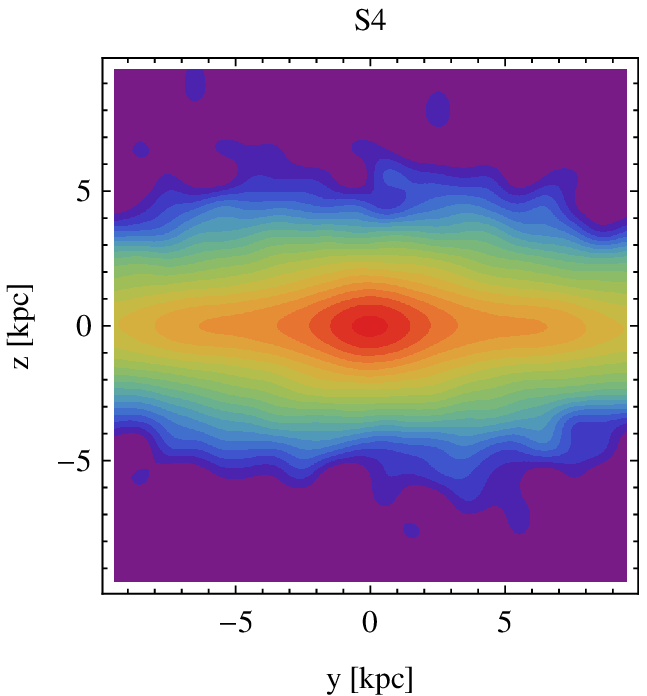}
\leavevmode
    \epsfxsize=0.805cm
    \epsfbox[53 -28 87 149]{legend.eps}
\leavevmode
    \epsfxsize=4.22cm
    \epsfbox[0 0 185 200]{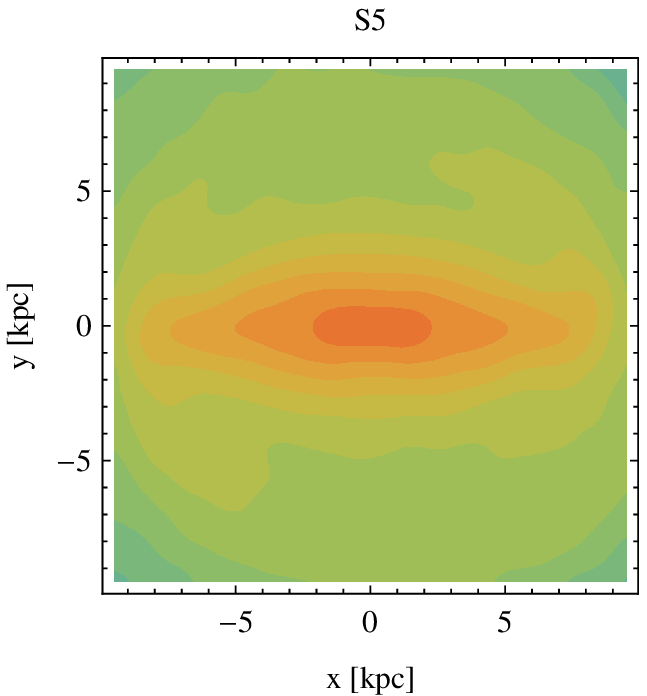}
\leavevmode
    \epsfxsize=4.22cm
    \epsfbox[0 0 185 200]{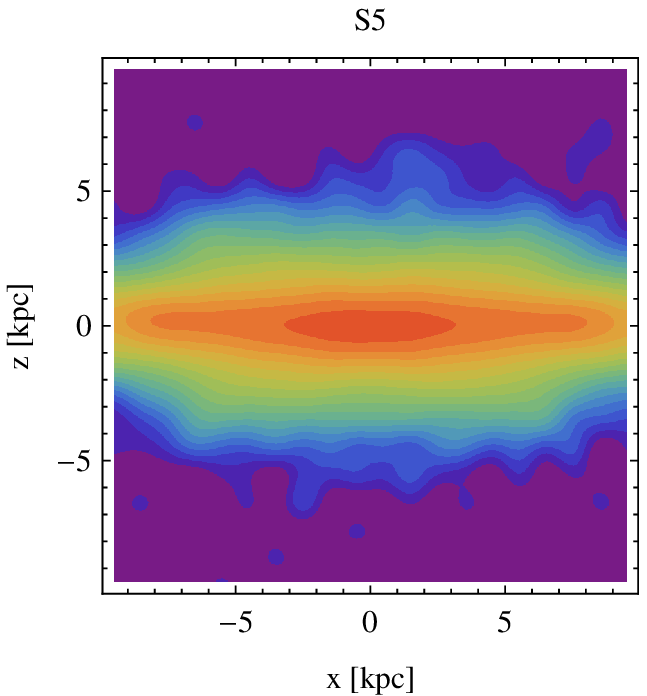}
\leavevmode
    \epsfxsize=4.22cm
    \epsfbox[0 0 185 200]{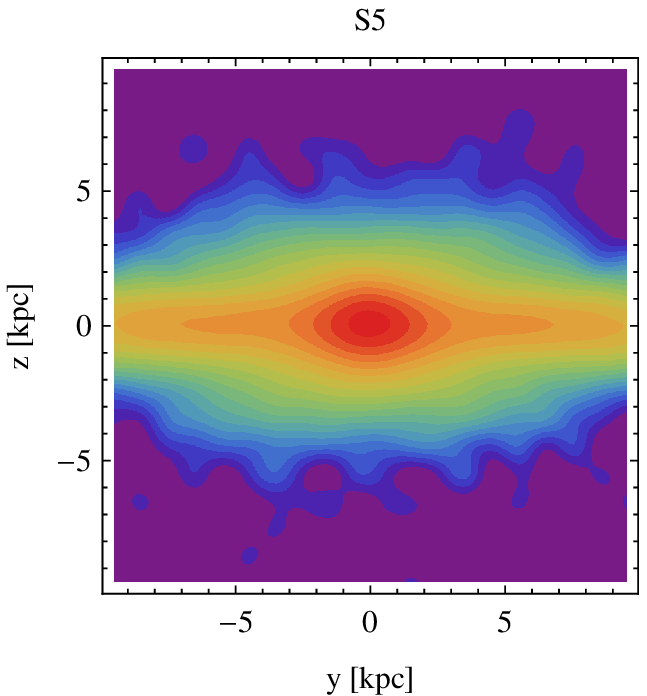}
\leavevmode
    \epsfxsize=0.805cm
    \epsfbox[53 -28 87 149]{legend.eps}
\end{center}
\caption{Surface density distributions of the stars in the simulated galaxies at the end of the evolution
for our five simulations (rows) and along different lines of sight: the shortest ($z$), intermediate ($y$)
and longest ($x$) axis of the stellar component (columns, from left to right). The surface density
measurements were normalized to log of maximum value $\Sigma_{\rm max} = 3.3 \times 10^9$ M$_\odot$ kpc$^{-2}$
occurring for the line of sight along the $x$ axis for S2. Contours are equally spaced in $\log \Sigma$ with
$\Delta \log \Sigma = 0.05$.}
\label{surdenrot}
\end{figure*}

The last
panel of the Figure shows the behavior of the anisotropy parameter of the stars $\beta$ (defined in the standard way as
$\beta = 1 - (\sigma_{\theta}^2 + \sigma_{\phi}^{'2})/(2 \sigma_{r}^2)$ where the second velocity moment
$\sigma_{\phi}^{'2} =\sigma_{\phi}^2 + v_{\phi}^2$ includes rotation).
For all simulations the anisotropy increases in time from the negative values
corresponding to tangential orbits (due to the initial rotation) toward more positive values corresponding to
more radial orbits. For S1 and S2, where rotation is strongly diminished, the anisotropy is mildly radial, while
for the remaining cases the orbits are close to isotropic or weakly tangential (S4) at the final stages of evolution.

\subsection{Evolution of shape}

Figure~\ref{shape7} describes the evolution of the shape of the stellar component within 7 kpc. The upper panel of the
Figure plots the intermediate to longest axis ratio ($b/a$), the middle panel the shortest to longest axis ratio ($c/a$)
and the lowest one the triaxiality parameter $T = [1-(b/a)^2]/[1-(c/a)^2]$. The evolution of $b/a$ shows strong
variability at the earlier phases of the simulations for tighter orbits S1-S3,
but the overall trend is for the $b/a$ to become lower in time. The evolution
of $c/a$ is much smoother: the thickness of the stellar component remains approximately
constant for the initial 1, 3, 4 and 6 Gyr
respectively for simulations S1, S2, S3 and S4/S5. After that the distribution of the stars becomes significantly thicker.
We discuss the origin for this behavior further below and relate it to the occurrence of the buckling instability.
The evolution of the triaxiality parameter from low values
characteristic of oblate disks towards high values characteristic of prolate spheroids strongly suggests
that in all
cases the progenitors experience the formation of a tidally induced bar. In the following section we discuss the
properties of these bars in more detail.

\section{Properties of the bars}

\subsection{Surface density distributions}

Figure~\ref{surdenrot} illustrates the final outcome of the simulations in terms of the surface density distribution
of the stellar component of the galaxies. The images were created by aligning the stellar components with the
principal axes determined as before from the distribution of stars within 7 kpc from the center.
The results for simulations S1-S5 are shown in rows. The
columns (from the left to the right) contain the projections along the shortest ($z$), intermediate ($y$)
and longest ($x$) axis of the stellar component, i.e. they correspond to the face-on, edge-on and end-on views of the
bars. The bar along the $x$ axis is clearly visible in all cases.

The shape of the bar is however significantly different in S1 and S2 (the two tightest orbits) than in the remaining
simulations: it shows a distinct boxy/peanut shape characteristic of a strong bar, not only in the edge-on but also
in the face-on view, while in S4 and S5
the isodensity contours are more elliptical, characteristic of weaker bars. Orbit S3 produces a bar
with a shape intermediate between the two. Note that while the surface density contours of the bars formed in
simulations S3-S5 would be well fitted by the generalized ellipses proposed by Athanassoula et al. (1990), the
shapes of bars in simulations S1-S2, especially in the outer parts, could be better approximated by curves similar to
Cassini ovals.

\subsection{Bar modes and pattern speeds}

\begin{figure}
\begin{center}
    \leavevmode
    \epsfxsize=8cm
    \epsfbox[0 10 186 186]{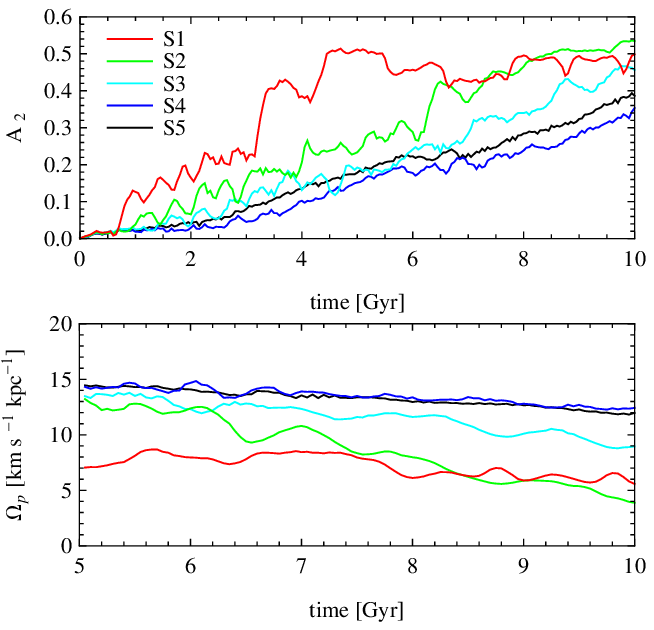}
\end{center}
\caption{Evolution of the bar mode $A_2$ measured for stars within the radius $r=7$ kpc (upper panel) and the
pattern speed of the bar (lower panel) as a function of time.}
\label{barmodepatternspeed7}
\end{figure}

\begin{figure}
\begin{center}
    \leavevmode
    \epsfxsize=7.8cm
    \epsfbox[0 10 180 180]{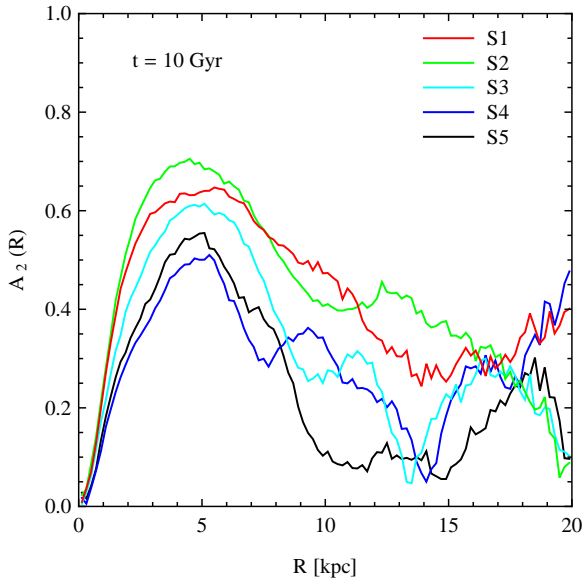}
\end{center}
\caption{The profiles of the bar mode $A_2 (R)$ at the final outputs of the simulations.}
\label{a2profiles}
\end{figure}

\begin{figure}
\begin{center}
    \leavevmode
    \epsfxsize=7.5cm
    \epsfbox[35 11.5 300 140]{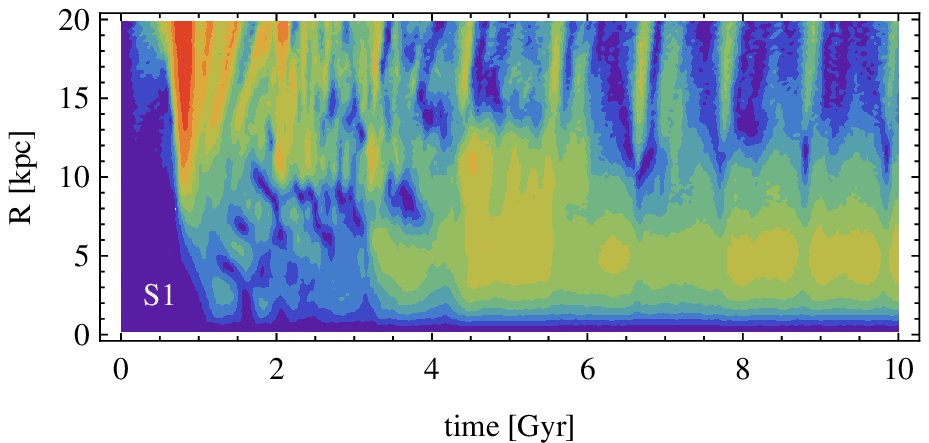}
\leavevmode
    \epsfxsize=0.837cm
    \epsfbox[38 -20 72 109]{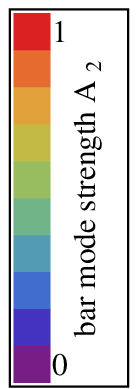}
\leavevmode
    \epsfxsize=7.5cm
    \epsfbox[35 11.5 300 140]{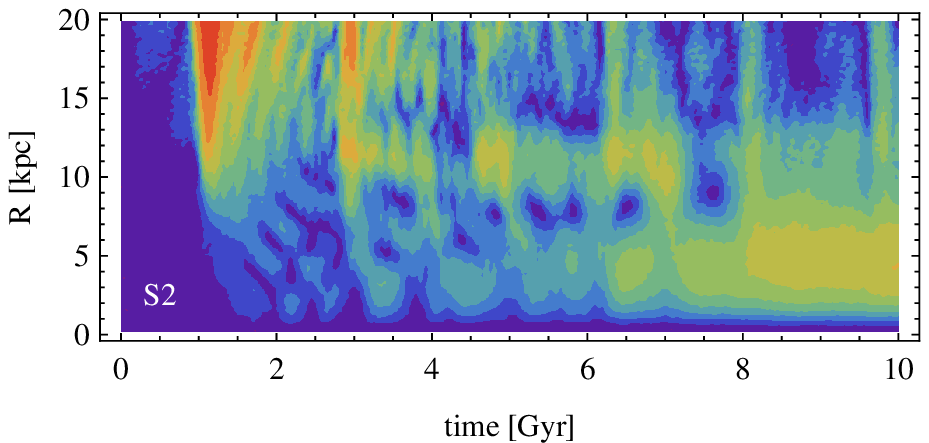}
\leavevmode
    \epsfxsize=0.837cm
    \epsfbox[38 -20 72 109]{legend2.eps}
\leavevmode
    \epsfxsize=7.5cm
    \epsfbox[35 11.5 300 140]{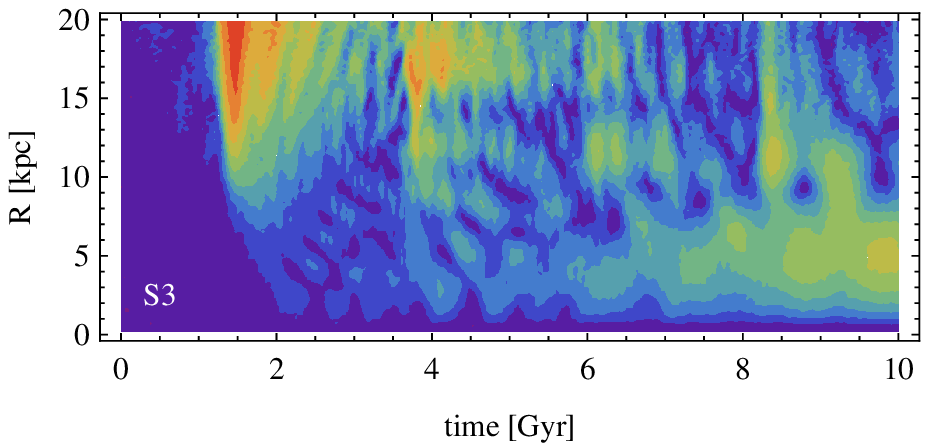}
\leavevmode
    \epsfxsize=0.837cm
    \epsfbox[38 -20 72 109]{legend2.eps}
\leavevmode
    \epsfxsize=7.5cm
    \epsfbox[35 11.5 300 140]{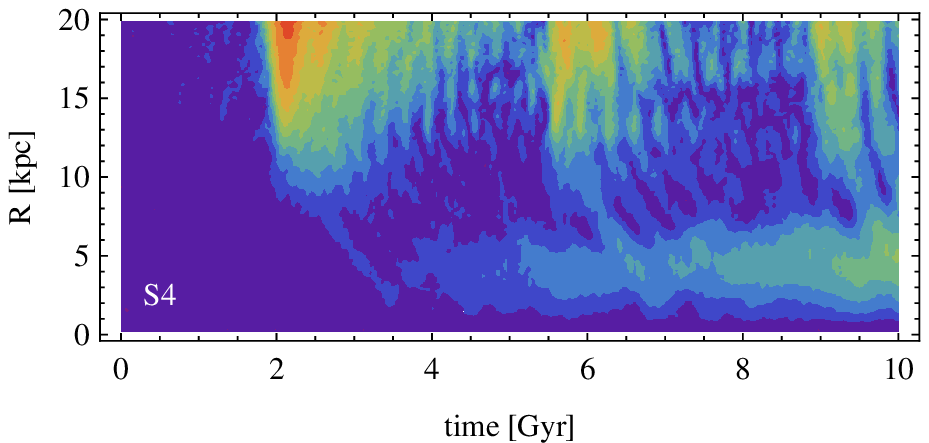}
\leavevmode
    \epsfxsize=0.837cm
    \epsfbox[38 -20 72 109]{legend2.eps}
\leavevmode
    \epsfxsize=7.5cm
    \epsfbox[35 11.5 300 140]{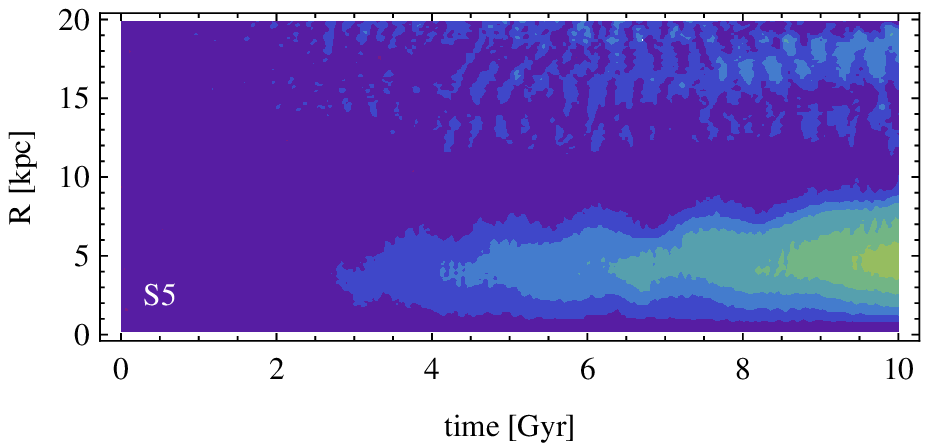}
\leavevmode
    \epsfxsize=0.837cm
    \epsfbox[38 -20 72 109]{legend2.eps}
\end{center}
\caption{The evolution of the bar mode profiles $A_2 (R)$ in time. The five panels from top to bottom show the results
for simulations S1-S5 respectively.}
\label{a2modestime}
\end{figure}

The strength of the bar can be characterized by one number, the module of the Fourier $m=2$ mode of the surface
distribution of the
stars. We calculated this parameter for each output of the five simulations projecting all stars along the shortest
axis and taking into account all stars within the radius of 7 kpc. The results are shown in the upper panel
of Figure~\ref{barmodepatternspeed7}. Confirming the impression from the surface density distributions discussed in
the previous subsection, we find the bar to be strongest in the case of simulations S1 and S2. Note however, that the
rather similar final stages of the two simulations are reached by significantly different paths. In the case of S1
the bar mode grows most strongly up to $A_2 = 0.5$ during the first 5 Gyr to remain more or less on the similar
level in the later stages of evolution.
On the other hand, for the remaining simulations S2-S5 this growth is more monotonic and stable, but at the end
of the evolution S2 even overcomes S1 reaching $A_2 \approx 0.54$.

In the lower panel of Figure~\ref{barmodepatternspeed7} we plot the pattern speed, i.e. the angular velocity of the bar,
measured during the last 5 Gyr, that is after the bar is formed in all simulations. As expected, the stronger is the bar
in terms of $A_2$, the lower its pattern speed. This correlation, known to exist for bars formed in isolation
(Athanassoula 2003)
is therefore confirmed also for our tidally induced bars. Again we find an approximately monotonic dependence of
the bar properties on the extent of the orbit of the galaxy: bigger tidal forces on tighter orbits produce
stronger bars with lower pattern speeds.

Interesting information about the bar properties can be obtained if we go beyond single-value measurements and
study the profiles of bar modes as a function of cylindrical radius in the galaxy, $A_2 (R)$.
Examples of such profiles for the final
outputs of the five simulations are shown in Figure~\ref{a2profiles}. The curves possess characteristic shapes,
with a maximum, typical of bars formed in isolation and similar to tidally induced bars in dwarfs
({\L}okas et al. 2014).

An even more complete characterization of the bar properties, previously applied e.g. by
Athanassoula et al. (2005) and Saha \& Maciejewski (2013), can be obtained by combining the dependence on distance
and on time to produce maps of $A_2$ mode profile evolution such as those shown in Figure~\ref{a2modestime}. Here
we combined the measurements like those shown in Figure~\ref{a2profiles} for 201 outputs of each simulation to create
contour plots with contours spaced by 0.1 in $A_2$. Such diagrams can be used to visually determine
both the bar strength and bar length depending on the preferred definition of
these quantities. For example, the first maximum value of $A_2$ along the radial variable can be adopted as a measure
of bar strength at a given time and some lower fixed value (e.g. half the maximum)
at a larger radius as a measure of the bar length (see the discussion in section 8 of Athanassoula \& Misiriotis 2002).

The diagrams allow us to grasp the full history of bar formation and evolution in the five simulations. Slightly after
the first pericenter passage that occurs at 0.7, 0.95, 1.25, 1.9 Gyr for simulations S1-S4 respectively, the disks are
stretched causing an increase of $A_2$ at large distances. As expected, the galaxy at the tightest orbit (S1) is affected
most strongly and down to lowest radii. For this case, the evolution is rather complicated: initially a small bar,
of radius about 3 kpc forms that only later grows to reach a typical scale of the order of 7 kpc and even larger.
Intermediate stages of this evolution involve the formation of spiral arms, rings and oval disks. Interestingly,
for this simulation, the bar appears longest and strongest between 4.5 and 5.5 Gyr, i.e. during one full orbit
between the fourth and fifth pericenter passage. After this period the bar is shortened and weakened until the
seventh pericenter at 7.7 Gyr when it starts to grow again. As discussed in section 3.2, this behavior is due to
particular orientations of the bar with respect to the tidal torque at pericenters. When the torque speeds up the
bar it makes it weaker, while the opposite is true for the torque slowing down the bar, which makes it stronger.
We note that at the end of evolution the bar in S1 is significantly longer than
7 kpc as confirmed by Figure~\ref{a2profiles} whether we estimate the length as the radius where the first minimum
occurs ($\sim 14$ kpc) or as the radius where $A_2$ falls down to half the maximum value ($\sim 12$ kpc).

For the remaining simulations on more extended orbits the situation is similar in a sense that a small bar
forms first and then grows to a size of at least 7 kpc but in general this happens later, in more steady manner and
the bar at the end is weaker. In all cases the disk is stretched at the first pericenter passage, but at larger
radii for more extended orbits. However, we always see perturbations (regions of slightly higher $A_2$)
propagating towards the center of the galaxy that seem to seed the bar. These perturbations are stronger and
propagate faster for tighter orbits.

The galaxy evolved in isolation (S5) remains stable against bar formation for the
first 3 Gyr. After this time a bar starts to develop and grows uniformly in a manner similar to the most extended
orbit (S4). However, the growth in S5 is slightly faster and the final bar is a little stronger and longer than for S4.
This behavior is probably due to the fact that, contrary to S5, in S4 the galaxy is subject to, however mild, tidal
force that affects and perturbs the outer parts of the disk forming tidal extensions. Only after that a perturbation
from the outer part travels towards the center to seed the bar.

\begin{table}
\begin{center}
\caption{Properties of the bars at the end of evolution}
\begin{tabular}{lccccc}
\hline
\hline
Simulation &   $A_{2,{\rm max}}$ & $a_{\rm b}$ & $\Omega_{\rm p}$ & $R_{\rm CR}$ & $R_{\rm CR}/a_{\rm b}$\\
           &                      &   [kpc]     & [km s$^{-1}$kpc$^{-1}$]       &  [kpc]    &    \\
\hline
\ \ \ \ S1  &  0.65     & 12.2      &\, 5.6   & 25.6  & 2.1 \\

\ \ \ \ S2  &  0.70     & 10.7      &\, 3.9   & 36.3  & 3.4 \\

\ \ \ \ S3  &  0.62     &\,  8.7    &\, 9.0   & 20.4  & 2.3 \\

\ \ \ \ S4  &  0.51     &\,  7.7    & 12.4    & 15.7  & 2.0 \\

\ \ \ \ S5  &  0.56     &\,  8.4    & 12.0    & 17.0  & 2.0 \\

\hline
\label{barproperties}
\end{tabular}
\end{center}
\end{table}
\begin{figure}
\begin{center}
    \leavevmode
    \epsfxsize=7.8cm
    \epsfbox[0 10 185 195]{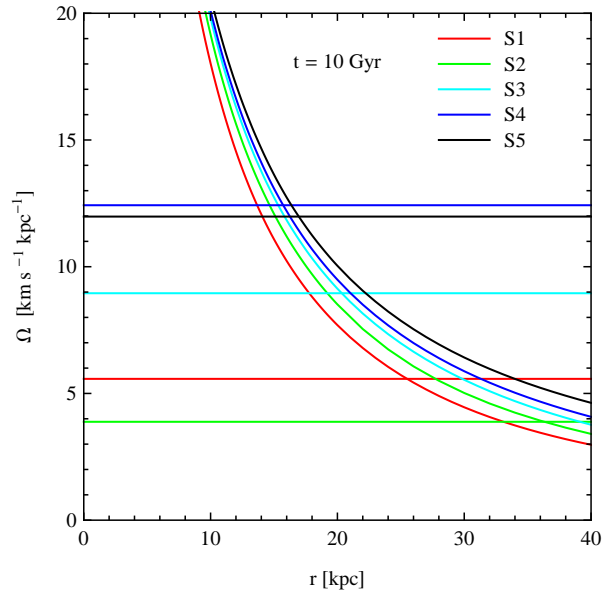}
\end{center}
\caption{The profiles of the circular frequency at the final outputs of the simulations (decreasing lines)
and the pattern speeds of the bars at the same time (horizontal lines).}
\label{corotation}
\end{figure}

Table~\ref{barproperties} summarizes the properties of the bars at the end of evolution. The columns of the Table
list (from the second to the sixth) the value of the maximum of the bar mode profile $A_{2,{\rm max}}$
(see Figure~\ref{a2profiles}), the length of the bar $a_{\rm b}$, the pattern speed $\Omega_{\rm p}$, the
corotation radius $R_{\rm CR}$ and the ratio $R_{\rm CR}/a_{\rm b}$. The corotation radii were estimated
from the comparison between the circular frequency and the pattern speeds of the bars in the final simulation outputs,
as shown in Figure~\ref{corotation}. Note that the mass loss due to tidal stripping is relatively mild in these
simulations as all curves are close to the original one (well approximated by the black line corresponding to the
galaxy evolved in isolation). In particular, even the galaxy on the tightest orbit (S1) still contains 40\% of the initial
mass within 40 kpc at the end of evolution. The length of the bar was estimated as the
cylindrical radius where the value of $A_2 (R)$ drops to $A_{2,{\rm max}}/2$ or, if a well-defined minimum of the profile
occurs at a smaller radius, this radius was adopted as the bar length. In most cases the ratio $R_{\rm CR}/a_{\rm b}$
is of the order of 2 indicating that our bars are slow. Our strongest bar in S2 is even slower with the ratio
equal to 3.4.

Another feature differentiating the evolution in the five simulations is the formation and survival of tidally induced
spiral arms. In all cases tidal extensions in the form of two-armed spirals are formed at pericenters.
In the case of S1 however the spiral arms are short-lived and are quickly dispersed forming extended tidal tails.
Once the stars in the outer parts are stripped no spiral arms are formed and the whole stellar component
is contained in the bar, as confirmed by low values of $A_2$ at $R > 13$ kpc
in the maps shown in Figure~\ref{a2modestime}.
In the case of S2-S4 the material in the spiral arms formed at the pericenter remains in the vicinity of the galaxy for
some time. The arms wind up to form tighter structures but survive for much longer, especially in the case of simulation
S4. We will discuss the properties of such
tidally induced spiral arms in more detail in a follow-up paper (see also Semczuk \& {\L}okas 2015).

\subsection{Buckling}

\begin{figure}
\begin{center}
    \leavevmode
    \epsfxsize=8cm
    \epsfbox[0 0 190 200]{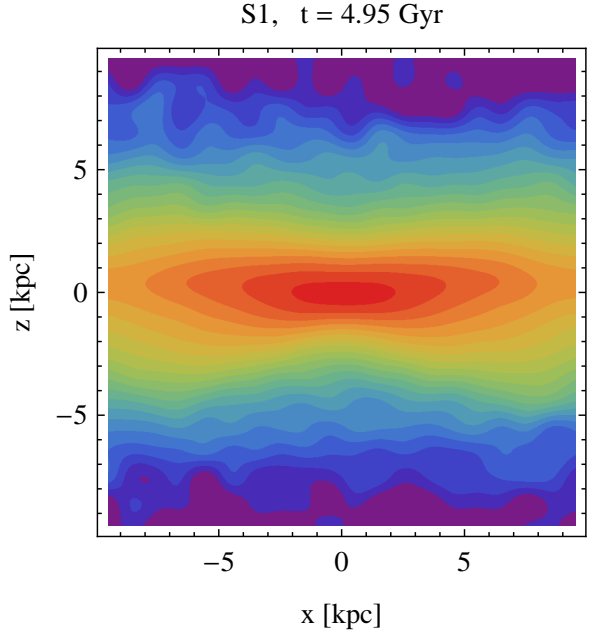}
\end{center}
\caption{An example of the buckling instability occurring for simulation S1 at 4.95 Gyr after the start of the simulation.
The plot shows the surface density distribution of the stars viewed edge-on, perpendicular to the bar major axis.
Asymmetries in the stellar distribution, characteristic of buckling, are clearly seen.}
\label{buckling}
\end{figure}

Another interesting phenomenon occurring during the evolution of the bars is that of buckling instability. An example
of the edge-on view of the bar in simulation S1 at 4.95 Gyr from the start clearly showing the distortion due to
buckling is presented in Figure~\ref{buckling}. One more example is seen in the middle panel for simulation S3 in
Figure~\ref{surdenrot}. It turns out that bars buckle in all our simulations, but later on in the evolution
for more extended orbits. We find that a good signature of the buckling is the presence of non-zero streaming velocity
along the shortest axis of the bar which is best measured using cylindrical coordinates.

\begin{figure}
\begin{center}
    \leavevmode
    \epsfxsize=8cm
    \epsfbox[0 10 186 186]{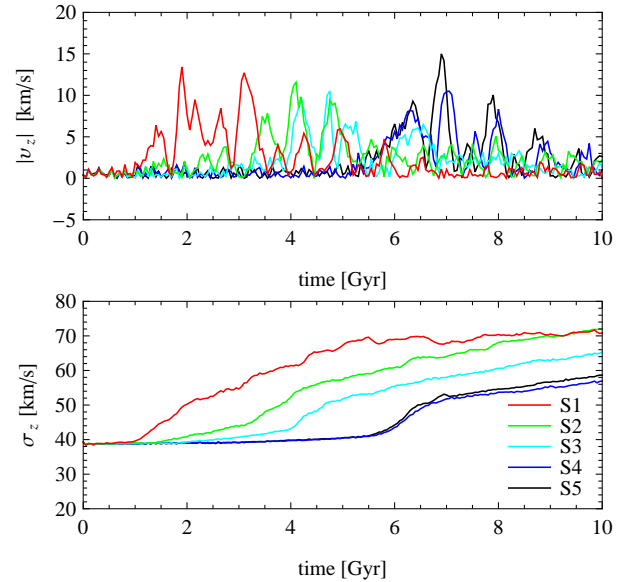}
\end{center}
\caption{The evolution of the mean velocity along the shortest axis and the corresponding velocity dispersion
as a function of time. The measurements were performed in cylindrical coordinates for stars within 7 kpc.}
\label{buckle7}
\end{figure}

The upper panel of Figure~\ref{buckle7} plots the absolute value of the mean streaming motion along the vertical
(shortest) axis $z$: $|v_z|$ as a function of time. All simulations show significant signal in this parameter,
up to about 15 km s$^{-1}$,
corresponding to distortions of the stellar distribution from the bar plane as present during the buckling instability.
This streaming motion occurs earliest in the evolution and is quite large for simulation
S1, but similar values of this velocity are measured also for the remaining simulations and at multiple occurrences.

Note that the first occurrence of large $|v_z|$ for simulation S1 happens almost immediately after the first small bar
is formed, i.e. around 1 Gyr after the start of the simulation. For simulations S2 and S3 the bars start to buckle
around 3 and 4 Gyr, again when the bar is still quite small and weak. In the case of S4 and S5 the bars start
to buckle much later, both around 6 Gyr, when the bars are longer and stronger. These timescales correspond
approximately to the second pericenter passage, at least for S2-S4, so the strong tidal force experienced at these
time may help to induce buckling. However, this interpretation is not supported by the fact that in the model evolved
in isolation (S5) buckling occurs at the same time as in S4.

In the lower panel of Figure~\ref{buckle7} we plot the evolution of the dispersion of the velocity in the vertical
direction, $\sigma_z$.
Clearly, the dispersion increases significantly at times corresponding to the presence of the significant
streaming motions in the upper panel of the Figure. Together with the increasing thickness of the stellar distribution,
as shown in terms of the ratio $c/a$ in the middle panel of Figure~\ref{shape7}, occurring at about the same time,
this means that the orbital structure of the bars experiences significant rebuilding. However, only for the
tightest orbits (S1 and S2) for which $c/a$ and $\sigma_z$ grow most strongly, the buckling results in the
formation of a clear boxy/peanut shape, well visible in the surface density plots of Figure~\ref{surdenrot}.
As expected from earlier studies (e.g. Martinez-Valpuesta \& Shlosman 2004)
the buckling episodes are accompanied by a slight weakening of the bar, or rather a slow-down in its
growth in terms of $A_2$, best visible in the upper panel of Figure~\ref{barmodepatternspeed7} for simulations S4 and S5.

We note that so far only mainly single or double buckling episodes were reported in bars studied in
the literature (Martinez-Valpuesta et al. 2006).
Even when recurrent, the first bucklings were found to be rather short-lived.
In contrast, in our bars (even the one formed in
isolation) buckling seems to last for an extended period, of the order of a few Gyr, affecting different parts of the bar.
It remains to be investigated if these phenomena are continuous or composed of multiple buckling episodes.

\subsection{Angular momentum transfer}

\begin{figure}
\begin{center}
    \leavevmode
    \epsfxsize=8.2cm
    \epsfbox[0 10 180 267]{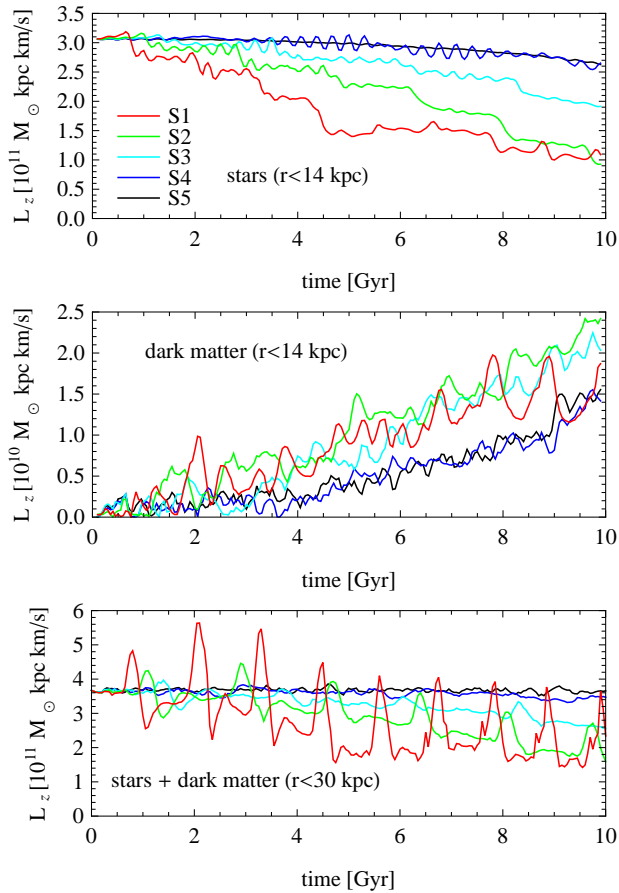}
\end{center}
\caption{Evolution of the $z$-component of angular momentum for stars (upper panel) and dark matter (middle panel)
for particles within the radius of 14 kpc from the center of the galaxy. The lower panel shows the sum of the
angular momentum for stars and dark matter but within 30 kpc.}
\label{angmomcomp}
\end{figure}

Transfer of angular momentum from the disk to the halo via the resonances
is believed to be the main mechanism behind the formation and evolution
of bars in isolation (Athanassoula 2003). Here we check whether a similar phenomenon
may be at work for tidally induced bars.
Figure~\ref{angmomcomp} plots the $z$-component of angular momentum of stars (upper panel) and dark matter (middle panel)
measured for particles within 14 kpc from the center of the galaxy as a function of time. Clearly, in general
the angular
momentum of stars decreases systematically with the most significant changes occurring at pericenter passages.

As expected, the behavior is qualitatively similar to the one of the rotation velocity shown in the upper panel of
Figure~\ref{kinematics7}. The smaller time-scale oscillations particularly visible in the case of S4 are due to
spiral arms. These oscillations are not present in S5, which indeed does not have transient spirals
(see Figure~\ref{a2modestime}). Note, however,
that on average S4 and S5 have nearly identical angular momentum evolutions, which indicates that, contrary to bars,
these transient spirals do not significantly contribute to the angular momentum exchange between the disk and the
halo.

The measurements for the dark matter particles show an opposite behavior. In this case the $z$-component of
angular momentum grows with time and more so for tighter orbits. For simulations S4 and S5 the amount of angular
momentum gained is almost identical, again pointing to the similarity between the two cases.
Note that the measurements for dark matter are much more noisy than for the stars because within this radius
the number of dark matter particles is about a factor of 10 lower than the number of stars.

In order to get a more complete picture, we have also measured angular momenta of stars and dark matter
for all particles within 30 kpc. The choice of this radius was motivated on one hand by the fact that almost all stars
are contained within this radius initially and on the other by this scale being of the order of the tidal radius
of the galaxy. The tidal radius depends on the orbit and time but its value (as estimated
using the prescriptions of Gajda \& {\L}okas 2016 for the prograde case) is close to 30 kpc
at the last pericenter of simulation S2 (smaller for S1, larger for S3 and S4).

The lower panel of Figure~\ref{angmomcomp} shows the sum of the $z$-components of angular momentum of stars
and dark matter within 30 kpc. For S5 (the galaxy evolved in isolation) the sum is constant in time, which means
that the amount of angular momentum lost by the stars is equal to the amount gained by dark matter.
This is also almost the case for S4 but for tighter orbits we see clear loss of the total angular momentum
over time, at least when measured between pericenters. At pericenters there are strong peaks from dark matter measurements,
especially for S1, but also for S2 and S3, because the dark matter component then experiences strong tidal torques.

This means that for tighter orbits some angular momentum is transferred outside the gravitationally bound body of the
evolving galaxy.
However, after 10 Gyr only 10\% of stars from within the 30 kpc radius are stripped even for the tightest orbit S1.
This small number of stripped stars is unable to carry away a significant amount of angular momentum. Therefore,
it must be transferred to the dark matter component first.
The dark matter halo, being more extended, is more heavily stripped by tidal
forces and the angular momentum can be carried away by its particles.

\section{Bar fraction in the cluster}

We have demonstrated that on average bars form faster and are stronger for tighter orbits. We may therefore expect
galactic bars to be stronger and more frequent near centers of clusters. Here we attempt to predict the
dependence of the bar fraction on the distance from the cluster center. For this purpose we
construct a toy model of the Virgo cluster using all the 800 simulation outputs we have available from our
four runs on different orbits S1-S4 (200 per
simulation, excluding the initial configurations). We therefore assume that for the last 10 Gyr
our toy Virgo cluster constantly accreted galaxies, four of them
every 0.05 Gyr, each on one of the four orbits we considered.
Combining the simulation outputs in this way, we obtain a sample of galaxies at a variety of distances from the
cluster center and at different evolutionary stages. We verified that the projected
density distribution of such a sample of
galaxies can be approximated by a power-law not very different from the actual distribution of galaxies in Virgo.

\begin{figure}
\begin{center}
    \leavevmode
    \epsfxsize=8cm
    \epsfbox[0 10 186 182]{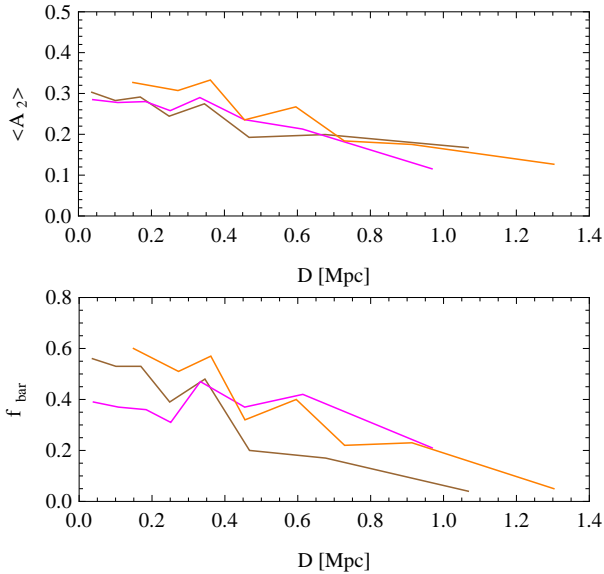}
\end{center}
\caption{The average value of the bar mode $A_2$ (upper panel) and the bar fraction (lower panel)
as functions of the projected distance from the cluster center. The bar fraction is defined
as the fraction of galaxies with $A_2>0.3$. In both panels the three lines correspond to
measurements along the three axes of the simulation box: $X$ (brown), $Y$ (magenta) and $Z$ (orange).}
\label{barfractionava2}
\end{figure}

We then assume that an imaginary observer is able to measure exactly the properties of the bars, in particular determine
the strength of the bar mode $A_2$ using stars within 7 kpc for all galaxies in the sample, as we have done in previous
sections. We then have a sample of 800 galaxies with known distances and $A_2$ values. We now translate the 3D
distances to the projected ones assuming that the cluster is observed along $X$, $Y$ and $Z$ axis of the simulation
box. Binning the galaxies in projected radius into 8 bins of 100 galaxies each, we calculate the mean value of $A_2$ and
the fraction of
galaxies with a strong bar ($A_2>0.3$) in each bin. The results for the quantities calculated in this way are shown
in Figure~\ref{barfractionava2} for the three lines of sight. The average strength of the bar (upper panel) shows
a clear trend of values decreasing with the distance from the cluster center. The bar fraction (lower panel)
also shows a general trend to decrease
with radius, although it varies strongly with distance and depends on the line of sight. The case of the  decreasing
bar fraction is least convincing
for the observation along the $Y$ axis (magenta line) where it is fairly constant with radius within the distance of
0.6 Mpc.

Given this rather idealistic realization of the Virgo cluster we expect that detecting such trends in real data
may be even more difficult. Recall that we have used just a single model of a progenitor galaxy and we used one initial
orientation of its disk with respect to the orbit (exactly prograde). Varying these parameters would result in a sample
of barred and un-barred galaxies whose average properties are very difficult to estimate. We have verified however by
running the simulations on orbits S2 and S3 with exactly
retrograde disk orientations that bars formed in such configurations
are similar to those in isolation, i.e. such disk orientations do not speed up the formation of bars. For the
mildly prograde and perpendicular orientations of the disk we expect some enhancement in the bar strength, but much
weaker than for the exactly prograde cases ({\L}okas et al. 2015). Therefore, if
galaxies are accreted by clusters with random disk orientations we may expect average bar strength and bar fractions
to decrease with respect to what we showed in Figure~\ref{barfractionava2}.

It would be essential to compare the predictions to real data. Unfortunately, although Virgo is one of the closest
and best studied clusters, no uniform morphological classification of its member galaxies has been performed to date,
although one may be available soon within the Next Generation Virgo Cluster Survey (Ferrarese et al. 2012). We have
therefore selected probable Virgo members from the NED database using the velocity criterion
$-1000$ km s$^{-1} < v < 3000$ km s$^{-1}$ that corresponds to $\sim 3 \sigma$ cut and removes obvious
interlopers. Such a selection yields
a sample of 1168 galaxies with morphological classification (barred vs. unbarred) up to 2.9 Mpc from the cluster center.
The morphological types given by the NED database were verified and in some cases supplemented using the data from LEDA
and Galaxy Zoo.

Calculation of
the fraction of barred galaxies as a function of radius gives values of the order $0.1 - 0.15$ almost independently
of radius out to 2 Mpc. There is thus no strong variation of the bar fraction visible in the data. Also the value found
is significantly lower than our prediction in the lower panel of Figure~\ref{barfractionava2}. However, if we take
into account that our toy model only includes disky galaxies and none of them undergoes a full transformation into
a spheroidal/elliptical object, the comparison would be more meaningful if we used only late type galaxies in Virgo.
Selecting only spirals and S0s with firm morphological classification (in the same velocity range) we end up with
a much smaller sample of 356 galaxies within 2.9 Mpc. The fraction of objects classified as barred among those turns out
to be much higher and varies in the range of $0.25 - 0.45$ depending on binning. This value agrees quite well with our
prediction for the central part of Virgo, but still no clear trend with radius is visible.

We have already remarked that the prediction may be an overestimate due to the narrow class of orbital configurations
considered. Also the data we have used are far from uniform and complete as the NED database is a compilation of
information originating from different studies and surveys.
Comparison with the data in the case of the Virgo cluster may also be hampered by the particular
properties of the cluster itself. First, the cluster is known to be non-spherical and departing from equilibrium.
However, the most important feature from our point of view may be the fact that it is composed of a few distinct
groups. It may very well be that the presence of such groups obscures or even destroys the dependence of the
bar fraction on the distance from the cluster center as the galaxies may have evolved in group rather than cluster
environment in the first place.

\section{Discussion}

We studied the formation and evolution of tidally induced bars in late-type galaxies similar to the Milky Way orbiting
in a Virgo-like cluster. We placed our progenitor galaxy on four different orbits in the cluster and also evolved it
in isolation as a reference case. Bars form in the galaxies in all our simulations and we find an approximately
monotonic dependence of the bar properties on the strength of tidal force experienced during the evolution. The bars
form earlier and are stronger and longer for galaxies more affected by the tidal force from the cluster. All our
bars experience extended periods of buckling instability, but again this occurs earlier on tighter orbits.
The formation time and properties of the bar in the model evolved in isolation are very similar to those in the
galaxy on the most extended orbit, so for this orbit no enhancement in bar strength was seen. We therefore conclude
that tidal interactions can trigger and influence bar formation in the cluster center but not in the outskirts.
Our results agree with those of Miwa \& Noguchi (1998) who estimated the range of tidal forces that should result in
strongly tidally modified bars. This range turns out to correspond to the values of tidal force experienced by our
galaxies on tighter orbits.

Our tidally induced bars turn out to be quite slow in terms of the ratio of
the corotation radius to the bar length which we find to be
of the order of 2 or higher while the typical observed values are of the order of unity
(i.e. most of the observed bars are fast).
They also appear slow when compared to the bar pattern speed of the Milky Way which is estimated
to be of the order of $\Omega_{\rm b} \approx 50$ km s$^{-1}$ kpc$^{-1}$ (Gerhard 2011)
while our fastest bars have $\Omega_{\rm b}$ below 15 in the same units.
While Miwa \& Noguchi (1998), who simulated the formation of tidal bars as a result of encounters between two galaxies,
also found their tidal bars to be rather slow, in our case the speed of the bar seems unrelated to the tidal
origin. Note that all our bars are similarly slow, even the one formed in isolation, so their speed is rather due to
the initial configuration. In addition, our bars are much longer and stronger than the real Milky Way bar so they must
have lower pattern speeds. The trend of stronger bars having lower pattern speed is
the same in our bars as in those formed in isolation.

Only for the tightest orbits does the tidal evolution lead to the formation of distinctive boxy/peanut shape at the end
of the simulations. Here we agree with Noguchi (1996) and Miwa \& Noguchi (1998) who observed that tidal
bars have different, more boxy shapes than bars formed in isolation. On the other hand, Athanassoula \&
Misiriotis (2002) found that the shapes of the bars can be different even if they form in isolation and
suggested that this is related to the fraction of dark matter in the galaxy center and the amount of angular
momentum transferred
from the disk to the halo. Since all our progenitors have the same structural properties initially, it seems that
the shape is indeed controlled by the amount of angular momentum transferred, whether it is caused by
the different initial structure or is induced by tides.

We verified that the mechanism behind the tidal bar formation in the configurations discussed here, i.e. a
normal-size galaxy evolving in a cluster environment,
is the transfer of angular momentum from stellar to dark matter particles of the galaxy. Only after the dark matter
halo is stripped by the tidal forces, is the angular momentum transferred outside the galaxy.
Note that this is significantly different from the case
of a dwarf galaxy orbiting the Milky Way discussed by {\L}okas et al. (2014) where the stellar angular momentum
was carried away by the stars stripped from the dwarf and feeding the tidal tails. Here very little stripping of the
stellar component takes place so a different mechanism must be at work. This mechanism, the transfer of angular
momentum from the stars to the dark matter halo, is thus the standard one that has been invoked as the process
responsible for bar formation in isolated galaxies (Athanassoula 2003; Martinez-Valpuesta et al. 2006;
Debattista et al. 2006).

Additional differences between the tidally
induced bars in dwarfs and in normal-size galaxies studied here include the fact that in the dwarf the buckling is
very weak and the bar structure is rather
simple while here the stellar component undergoes a number of rapid variations, including extended periods of buckling,
and the formation of the bar is accompanied by the presence of short-lived
substructures like rings, spiral arms or even bars embedded within weaker bars or oval disks.
This is especially the case for the tighter orbits. On the other hand, on more extended orbits we witness
the formation of strong and long-lived spiral arms.
The observational properties of these tidally induced bars and spiral arms will be discussed
in more detail in follow-up papers.

The problem of the evolution of disky galaxies in cluster-like environments has been addressed in a number of past
studies. For example, Mastropietro et al. (2005) studied the evolution of disky dwarf galaxies and their transformation
into early type objects. They reported the formation of tidally induced bars in some of their configurations.
On the other hand, Gnedin (2003) and Smith et al. (2015) did not get bars as a result of tidal interactions in their
studies. The difference in the outcome of their simulations and ours can be traced to
different initial configurations: their disks are rather hot and thick from the start
(in the case of Smith et al. 2015) or become such when evolved in isolation (Gnedin 2003) and thus are more stable
against bar formation. In contrast, in our case the thickness and the vertical velocity dispersion of the stellar
component remains constant in time until the bar buckles.

Using all our simulation outputs we constructed a toy model of the population of galaxies in a Virgo-like cluster
in order to estimate the expected average strength of the bar and the fraction of barred galaxies
as a function of clustercentric distance. We predict both quantities to be mildly decreasing functions of radius.
We have also attempted a preliminary comparison with the data using a morphologically classified sample of galaxies
in Virgo from NED but were unable to confirm the expected trends. The difficulty lies in the lack of a proper
data set with uniform morphological classification and depth but also in the presence of substructure in Virgo that may
obscure the trends.

An obvious caveat of the results presented here is the lack of dissipational component in our simulations,
as including gas dynamics and star formation is known to have some influence on bar formation in galaxies
(Debattista et al. 2006; Athanassoula et al. 2013).
In particular, Athanassoula et al. (2013) find that in the presence of a significant gas fraction in the disk
bars form later and are weaker. However, the gas component in normal-size galaxies is probably not dominant at the time
they are accreted by a cluster and in addition the gas is expected to be stripped by ram
pressure rather quickly after the galaxies are accreted if they plunge deep enough inside the cluster,
as is the case for our orbital configurations (Quilis et al. 2000).
Therefore, in less that one orbital time the dynamics of the galaxies is expected to be essentially collisionless.
We are thus confident that our simplified approach is able to grasp the
basic evolution of disky galaxies in clusters and the process of formation of tidally induced bars in them.

\section*{Acknowledgments}

This work was supported in part by the Polish National Science Centre under
grant 2013/10/A/ST9/00023 and by the project RVO:67985815.
IE and AS acknowledge the hospitality of the Copernicus Center in Warsaw during
their visits. We thank L. Widrow for providing procedures to generate $N$-body realizations for initial conditions.
Helpful comments from an anonymous referee are kindly appreciated.
This research has made use of the NASA/IPAC Extragalactic Database (NED) which is operated by the Jet Propulsion
Laboratory, California Institute of Technology, under contract with the National Aeronautics and Space Administration.
We also acknowledge the use of LEDA and Galaxy Zoo databases.

\end{document}